\documentclass[%
 reprint,
 amsmath,amssymb,
 aps,
floatfix,
]{revtex4-2}

\usepackage{graphicx}
\usepackage{dcolumn}
\usepackage{bm}

\usepackage{color}
\usepackage{physics}
\usepackage[normalem]{ulem}

\begin{document}


\title{Phase dynamics in an AC driven multiterminal Josephson junction analogue}

\author{Fran\c{c}ois Amet$^{1,\ast}$, Sara Idris$^{1}$, Aeron McConnell$^{1}$, Brian Opatosky$^{1}$, Ethan Arnault$^{2}$
\\
\normalsize{$^{1}$Department of Physics and Astronomy, Appalachian State University, Boone, NC 28607, USA,}
\\
\normalsize{$^{2}$Department of Physics, Duke University, Durham, NC 27708, USA}
\\
\normalsize{$^\ast$To whom correspondence should be addressed; E-mail:  ametf@appstate.edu}
}

\date{\today}

\begin{abstract}
In the presence of an AC drive, multiterminal Josephson junctions exhibit the inverse AC Josephson effect, where the oscillations of the superconducting phase of each junction can lock onto one another or onto the external drive. The competition between these different phase locked states results in a complex array of quantized voltage plateaus whose stability strongly depend on the circuit parameters of the shunted junctions. This phase diagram cannot be explored with low temperature transport experiments alone, given the breadth of the parameter space, so we present an easily tunable analog circuit whose dynamical properties emulate those of a three terminal junction. We focus on the observation of the multiterminal inverse AC Josephson effect, and we discuss how to identify Shapiro steps associated with each of the three junctions as well as their quartet states. We only observe integer phase locked states in strongly overdamped networks, but fractional Shapiro steps appear as well when the quality factor of the junctions increases. Finally, we discuss the role of transverse coupling in the synchronization of the junctions. 

\end{abstract}

\maketitle

\section{Introduction}

In a multiterminal junction, a Josephson coupling is established between multiple superconducting electrodes across a common normal channel. Those devices recently attracted considerable interest\cite{Strambini2016, Cohen2018, Draelos2019, Arnault2021,Pankratova2020, Pribiag2020, KoFan2021} because of the energy spectrum of their Andreev bound states. Indeed these states, which result from the Andreev reflections of charge carriers at each superconducting interface, have an energy spectrum that can emulate artificial band structures with interesting topological features such as Weyl points and non-Abelian monopoles \cite{Riwar2016, Eriksson2017, Meyer2017, Xie2017, Xie2018, WeylCircuit, MelinBerry,MelinFloq}.  

While the quantum effects in these devices provide unique opportunities, the dynamical effects of these circuits can give rise to unexpected effects \cite{StrogatzSynch, Chimera, Splay1, Splay2}. This is because the time-evolution of the superconducting phases obey nonlinear differential equations that are comparable to those of driven coupled pendulums\cite{Tinkham, McCumber1968, Stewart1968,arnault_dynamical_2022,KautzRev1996}. As a result, many features that are traditionally recognized as quantum effects in Josephson junctions, may actually be caused by the nonlinearity of the equations describing them. For instance, fractional Shapiro steps, which are often attributed to the non-sinusoidal current phase relation (CPR) of a device, can actually be observed in two or three terminal junctions with a strictly sinusoidal CPR as a consequence of the classical equations governing them\cite{Arnault2021,Amet2021}. Another example arises in three terminal devices, where supercurrent resonances can occur when commensurate finite voltages are applied to each terminal. While those have been attributed to Andreev multiplets entangling four or more electrons, they can in fact have a strictly dynamical interpretation which is observable in classical systems\cite{arnault_dynamical_2022, Melo2021}. Therefore, as a way to distinguish the quantum realm from the classical, there is a need to realize an experimental platform that mimics the dynamical processes in these devices, decoupled from any complicating quantum effects. 

In the case of single junctions, analog circuits based on conventional components can be designed so that the voltage at one of their nodes follows the same differential equation as the junction phase in the RCSJ model\cite{magerlein_accurate_1978,hamilton_analog_1972,dhumieres_chaotic_1982,blackburn_circuit_2007,Amet2021}. Recently, we showed how a broad range of RCSJ phenomena could be observed in such a circuit\cite{Amet2021}. These include hysteretic switching, activated escape rate caused by thermal-noise, phase-locking and chaos. Further, the design allowed the observation of the time evolution of the phase itself, as well as its frequency spectrum. These are not experimentally observable in standard junctions, because the phase dynamics occur on sub-nanosecond timescales.  

In this work, we expand these results to the case of a three-terminal junction. We designed a circuit based on voltage-controlled oscillators (VCO) that follows the same dynamical system as a network of three shunted junction. The transport properties of this circuit not only replicate recent observations made on ballistic multiterminal junctions, but also unravels the phase dynamics of periodically driven devices. We observe the multiterminal inverse AC Josephson effect in this circuit, and explore how the stability of the resulting integer and fractional Shapiro steps depends on shunting parameters.

\begin{figure*}
    \centering
    \includegraphics[width=\textwidth]{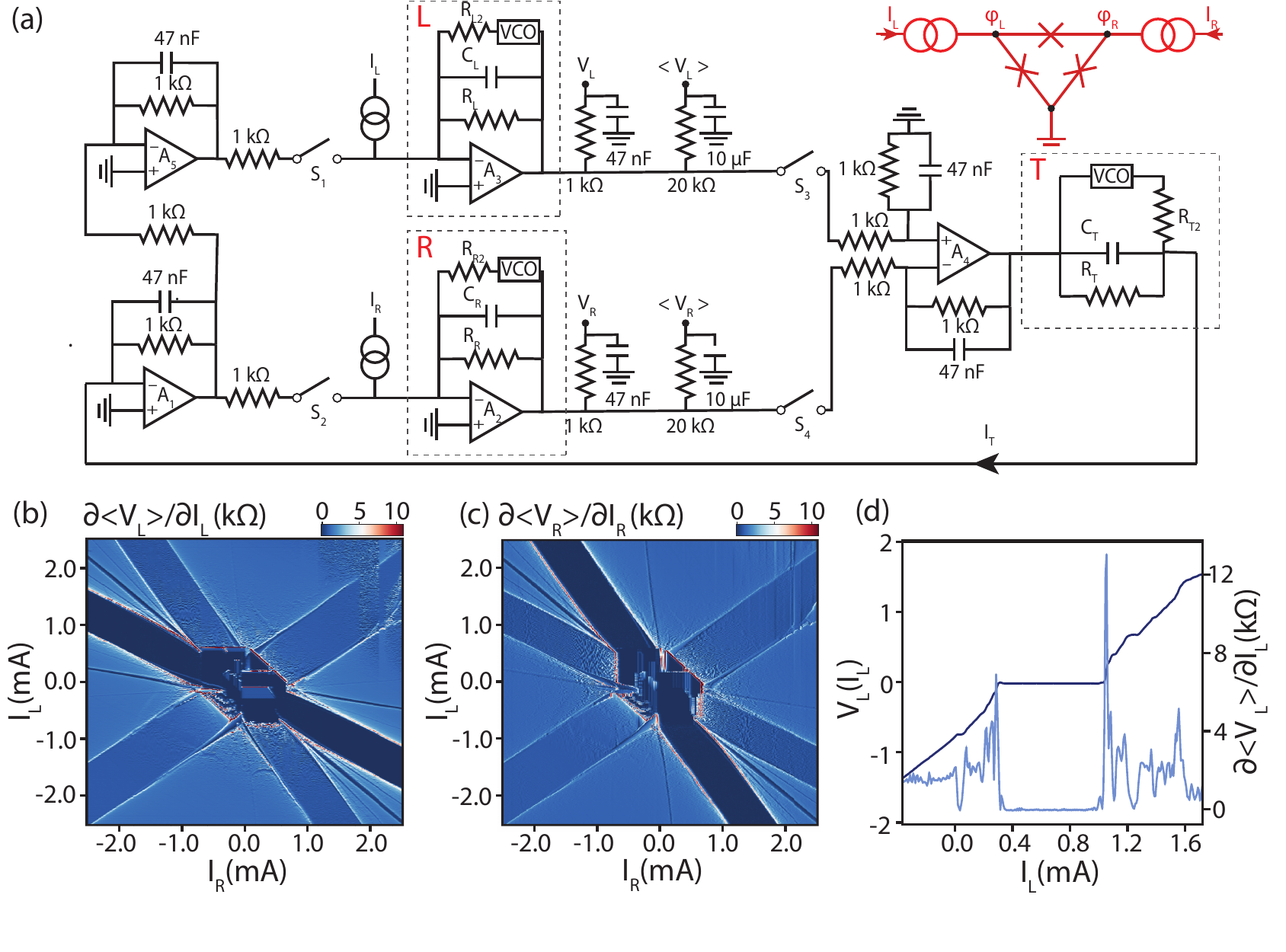}
    \caption{a) Schematic of the analog circuit used in this work. In red: schematic of an equivalent three terminal network of Josephson junctions. The bottom superconducting contact is grounded and its phase set as zero. Subcircuits corresponding to the left (L), right (T), and transverse junctions (T) are highlighted with dashed boxes. b) Effective differential resistance of the circuit $\partial \langle V_{L}\rangle/\partial I_{L}$ as a function of the two biases $I_{L}$ and $I_{R}$. c) Effective differential resistance of the circuit $\partial \langle V_{R}\rangle/\partial I_{R}$ as a function of the two biases $I_{L}$ and $I_{R}$. d) Cross section of $\langle V\rangle_{L}(I_{L})$ for $I_{R}$=-1.47 mA}
    \label{fig:1}
\end{figure*}

\section{The three terminal shunted junction model}

We model the three-terminal Josephson junction by the network which is sketched in red on Figure 1a. The three terminals are labeled L, R and B. The bottom contact is grounded, so its  phase is assumed to be 0. Each junction is assumed to be shunted by a resistor and a capacitor. Applying Kirchhoff laws at each node, we find\cite{suppinfo}:

\begin{equation}
\frac{\hbar}{2e}\mathcal{C} \ddot\Phi +\frac{\hbar}{2e}\mathcal{G}\dot\Phi + I_c(\Phi)=I
\end{equation}

Here, $\Phi=\begin{pmatrix} \varphi_{L} \\ \varphi_{R} \end{pmatrix}$ and $I=\begin{pmatrix} I_{L} \\ I_{R} \end{pmatrix}$ are two-row vectors, and we defined the following quantities:

\begin{align*}
I_{c}(\Phi)&=\begin{pmatrix}I_{L}\sin(\varphi_{L})+I_{T}\sin(\varphi_{L}-\varphi_{R})\\
    I_{R}\sin(\varphi_{R})+I_{T}\sin(\varphi_{R}-\varphi_{L})
    \end{pmatrix}
    \\
\mathcal{C}&=\begin{pmatrix}C_{L}+C_{T} & -C_{T}\\-C_{T}& C_{L}+C_{T}\end{pmatrix}
\\
\mathcal{G}&=\begin{pmatrix}G_{L}+G_{T} & -G_{T}\\
    -G_{T}& G_{L}+G_{T}
    \end{pmatrix}
    \\
\end{align*}
$I_c(\Phi)$ depends on the current phase relations of the junctions, which for simplicity are assumed to be sinusoidal. 
\begin{figure*}
    \centering
    \includegraphics[width=\textwidth]{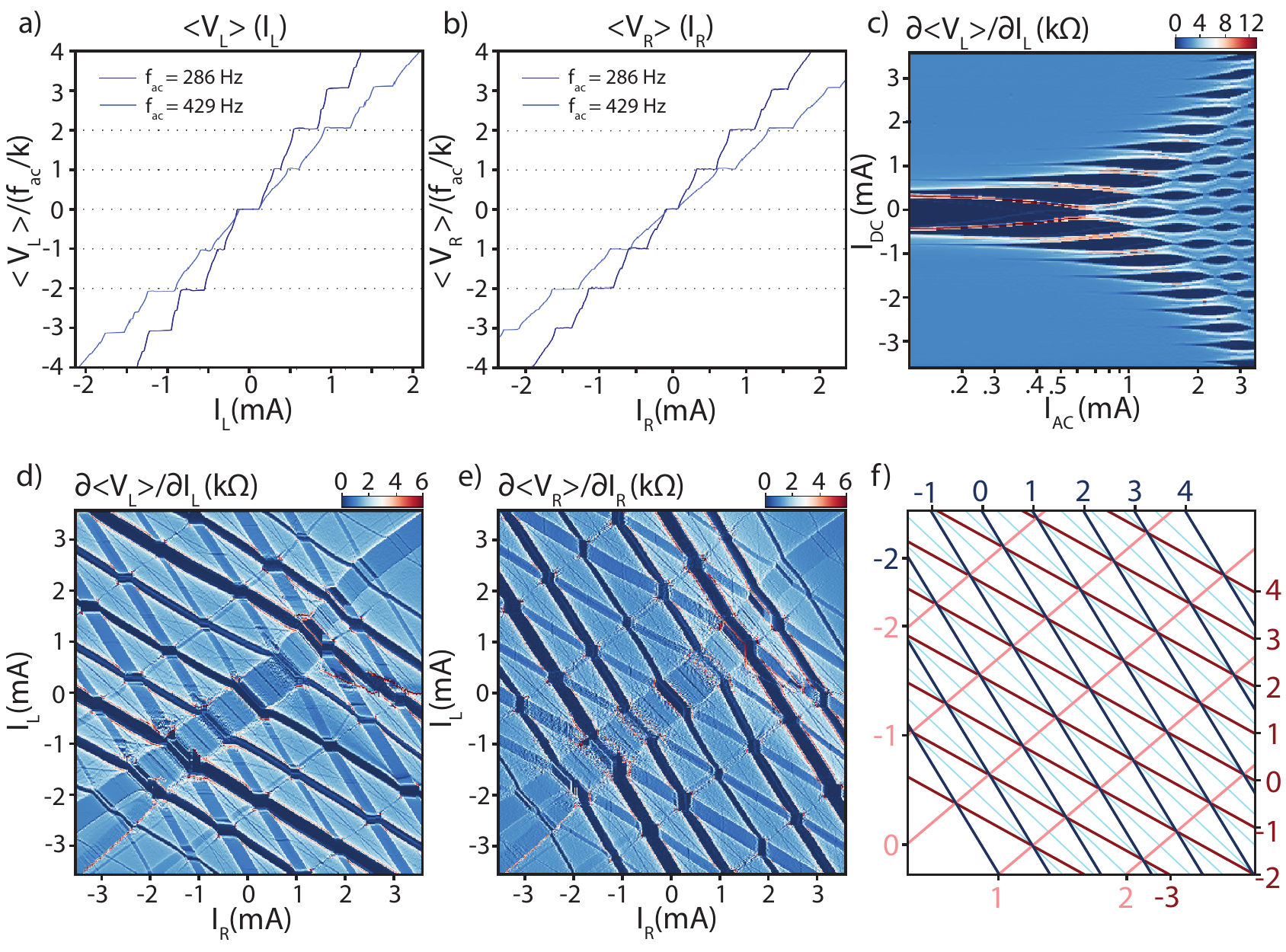}
    \caption{a) Shapiro steps in an AC driven uncoupled analog junction. $\langle V_{L}\rangle$ is plotted as a function of bias when $I_{ac}$=1.84 mA and $f_{ac}=429.2 Hz$ (light blue), or $I_{ac}$=1.13 mA and $f_{ac}=286.7 Hz$. The junction voltage is expressed in units of the voltage quantum $f_{ac}/k$, which is why the slope of the trace changes despite the unchanged quality factor. b) Shapiro steps for  $\langle V_{R}\rangle$ using the same parameters. c) Differential resistance $\partial\langle V_{L}\rangle/\partial I_{L}$ in the absence of coupling as a function of DC and AC biases at $f_{ac}=286.7 Hz$. d) Shapiro steps in the presence of transverse coupling. Differential resistance $\partial\langle V_{L}\rangle/\partial I_{L}$ as a function of both biases $V_{L}$ and $V_{R}$. The map was measured with $f_{ac}=429.2 Hz$ and $I_{ac}$=1.76 mA. e) Differential resistance of the other channel $\partial\langle V_{R}\rangle/\partial I_{R}$ as a function of both biases $I_{L}$ and $I_{R}$. f) Schematic of the main resonances to notice in Panels d and e. Integer plateaus in $\langle V_{L}\rangle$ are labeled in dark red, plateaus in $\langle V_{R}\rangle$ are labeled in dark blue, plateaus in the transverse junction voltage $\langle V_{L}\rangle/-\langle V_{R}\rangle$ are labeled in pink, and quartet plateaus in $\langle V_{L}\rangle+\langle V_{R}\rangle$ are labeled in light blue.}
    \label{fig:2}
\end{figure*}

The circuit shown in black on Figure 1a reproduces the same system of differential equations as the three terminal network of Josephson junctions. A simpler circuit analogue of a two terminal Josephson junction was already studied in Ref.\cite{Amet2021} and is here generalized to the three terminal case. It relies on three home-made voltage controlled oscillators (VCO) which have been simplified as box diagrams for clarity but are shown in the supplementary information. These oscillators deliver an output sine-wave of amplitude $\alpha$ and frequency $kV$, where $V$ is the input voltage of the VCO and $k$ its voltage to frequency gain. 

If we define $\dot{\varphi}_{L}$ and $\dot{\varphi}_{R}$ as the output voltages of operation amplifiers A3 and A2 (multiplied by $2\pi k$), we show in Ref.\cite{suppinfo} that $\Phi$ verifies the equivalent differential equation:
\begin{align}
    \frac{1}{2\pi k}\mathcal{C} \ddot\Phi +\frac{1}{2\pi k}\mathcal{G} \dot\Phi+I_{c}(\Phi)=I
\end{align}

It is thus possible to monitor the time evolution of output voltages of A2 and A3, and get insights into the phase-dynamics of three terminal Josephson junctions. Here, we defined the two vector $I$ as: $I =\begin{pmatrix}-I_{L}\\
-I_{R}\end{pmatrix}$.

\section{DC transport characterization}

We first evaluate the switching properties of the analog JJ network in the presence of a DC bias. Figures 1b and 1c show the effective differential resistances $\partial \langle V_{L}\rangle/\partial I_{L}$ and $\partial \langle V_{R}\rangle/\partial I_{R}$ as a function of DC biases $I_{L}$ and $I_{R}$. 

The maps draw strong similarities with previous work on three terminal junctions shown in Ref.\cite{Draelos2019,Pankratova2020,Pribiag2020}. Three arms of suppressed differential resistance correspond to each of the three junctions being in the zero-voltage state. The strongest resonances of suppressed differential resistance correspond to $V_{L}$=0, $V_{R}=0$ and $V_{L}-V_{R}$. The maps of those voltages are shown in Ref.\cite{suppinfo}.

A cross section of the differential resistance $\partial \langle V_{L}\rangle/\partial I_{L}$ as a function of $I_{L}$ is shown in light blue in Figure 1d. The corresponding I-V curve $V_{L}(I_{L})$ is shown in darker blue on the same plot. Those cross-sections are strongly reminiscent of typical Josephson junction transport, and evidently the region of suppressed differential resistance correspond to a plateau at $V_{L}=0$.

In addition to the three main zero-voltage resonances, the maps 1b and 1c show additional resonances along contours defined as $p\langle V_{L}\rangle+q\langle V_{R}\rangle=0$ with p, q $\in\mathbb{N}$. Those resonances are also seen on the cross sections 1d around $I_{L}\approx 0$ and $I_{L}\approx$ 1.25 mA. 
These correspond to classical realization of Andreev multiplet-states and were the focus of a different publication \cite{arnault_dynamical_2022}. 

\section{Integer phase locking}

We now turn to the transport properties of the network in the presence of an AC drive. Conventional multiterminal Josephson junctions can exhibit the multiterminal inverse AC Josephson effect in the presence of microwave radiation\cite{Arnault2021}. In the presence of a periodic drive, the junction phase can lock onto a multiple of the drive frequency $n\omega$\cite{Shapiro1963}, which results in a quantized voltage across the junction. This is a purely dynamical effect, which can thus be replicated in a Josephson junction analogue, as shown in Ref.\cite{Amet2021}. Here, the three analog junction frequencies are on the order of 500 Hz. It is therefore possible to observe phase-locking in the presence of an AC excitation provided by a simple function generator. 
\begin{figure*}
    \centering
    \includegraphics[width=\textwidth]{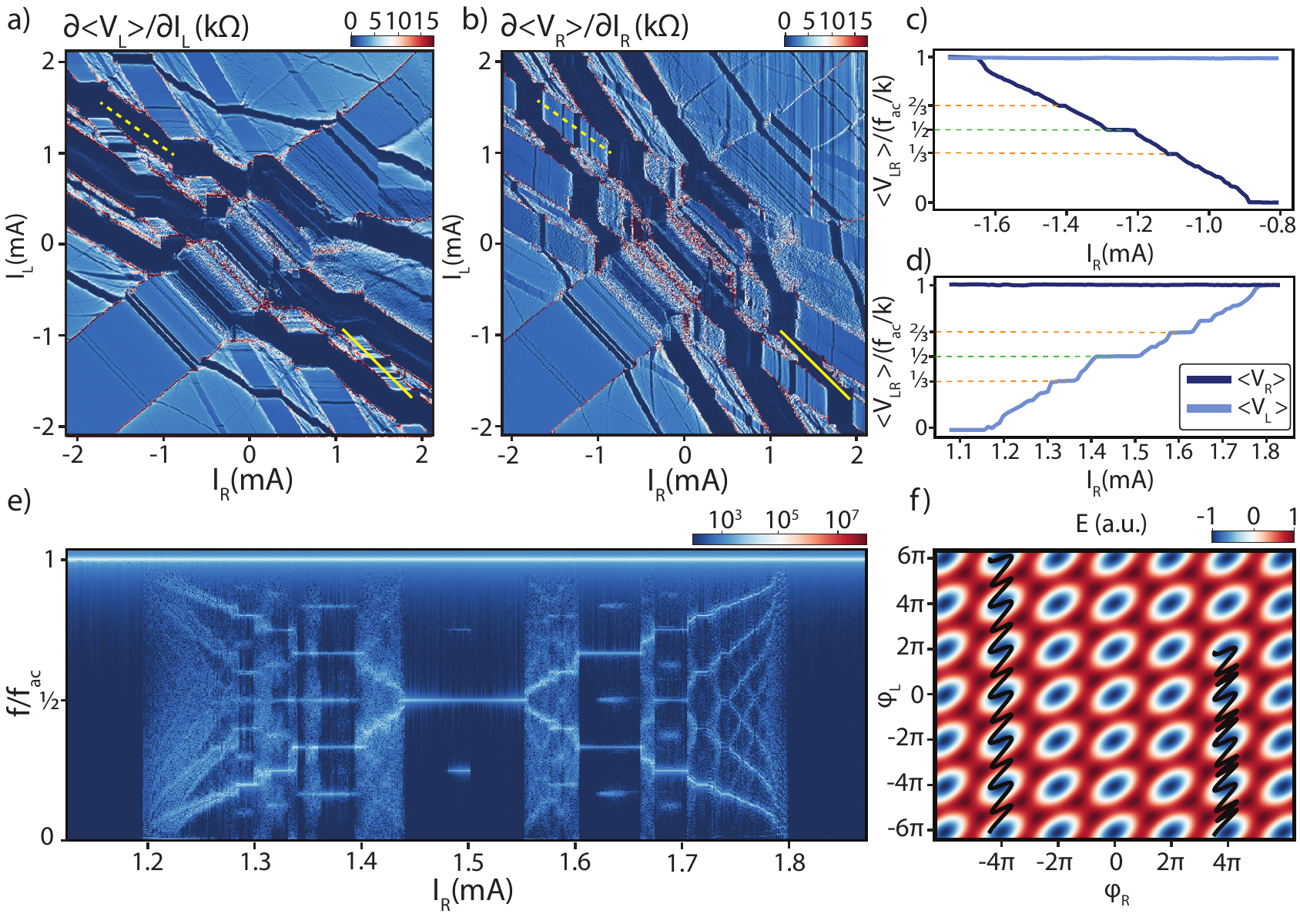}
    \caption{AC Fractional Shapiro steps in an driven analog junction network. a) Differential resistance $\partial\langle V_{L}\rangle/\partial I_{L}$ as a function of both biases $I_{L}$ and $I_{R}$. The map was measured with $f_{ac}=687.7 Hz$ and $V_{ac}$=1.2 mA. b) Differential resistance of the other channel $\partial\langle V_{R}\rangle/\partial I_{R}$. c) Cross section $\langle V_{L}\rangle$ and $\langle V_{R}\rangle$ through a fractional Shapiro steps sequence as a function of $I_{R}$. The corresponding cross section is shown in yellow dashes on panel a. It corresponds to an integer plateau of $\langle V_{L}\rangle$ which coincides with a sequence of fractional steps in $\langle V_{R}\rangle$. d) Similar cross section, corresponding to the continuous yellow line in panel a, where $\langle V_{R}\rangle$ is quantized and $\langle V_{L}\rangle$ goes through a sequence of fractional steps. e) Frequency spectrum of $V_{L}(t)$ as a function of bias along the cross section shown in panel a by the bold yellow line. f) Potential profile of the three terminal junction. Two simulated trajectories in phase space are overlaid on top of the map, and each correspond to 12 cycles of the drive. The trajectory on the left corresponds to plateaus $\langle V_{R}\rangle$=0 and $\langle V_{L}\rangle=\frac{1}{2}\frac{f_{ac}}{k}$, while the one on the right corresponds to $\langle  V_{R}\rangle$=0 and $\langle  V_{L}\rangle=\frac{1}{3}\frac{f_{ac}}{k}$.  }
    \label{fig:3}
\end{figure*}

We first characterize the inverse AC Josephson effect when the coupling between the Josephson junctions is disconnected (using the switches shown in Figure 1a). Here the three analog junctions have a quality factor of about 0.3 and are thus overdamped. 
We observe Shapiro steps in $\langle V_{L}\rangle(I_{L})$ and $\langle V_{R}\rangle(I_{R})$, shown on panels 2a and 2b. The voltage quanta are $f_{ac}/k$, where k is the voltage to frequency gain of the VCO. 
Similar to the conventional Josephson junction case, the height of the voltage steps is expected to be proportional to the AC drive frequency $f_{ac}$. For each junction, we thus plot the normalized output voltage at two different frequencies, and find that the quantization indeed scales with $f_{ac}$. 

We can then determine the evolution of the width of the Shapiro steps as a function of the amplitude of the AC drive. In the case of a voltage biased Josephson junction, that width can be determined analytically and follows Bessel-like oscillations as a function of the drive amplitude. While this is not the case in current-biased junctions, oscillations are still observed and can be perfectly replicated within the RCSJ model\cite{Amet2021}. Figure 2c shows that trend: dark blue regions correspond to a vanishing  differential resistance $\partial \langle V_{L}\rangle/\partial I_{L}$ and therefore quantized Shapiro steps in the I-V curve of the junction. This map is typical of overdamped behavior, as observed in both standard junctions\cite{Trevyn} and analog junctions\cite{Amet2021}. 

Now that signatures of phase-locking in uncoupled analog junctions are established, we restore the transverse coupling and determine the evolution of the differential resistances $\partial \langle V_{L}\rangle/\partial I_{L}$ and $\partial \langle V_{R}\rangle/\partial I_{R}$ as a function of both biases (Figure 2d and 2e). Important patterns in the data are sketched in Figure 2f. Shapiro plateaus are observed in both channels and they correspond to the darkest blue stripes observed on Figure 2d and Figure 2e. Plateaus of constant $\langle V_{L}\rangle$ in Figure 2d are sketched as dark red lines in 2f, whereas plateaus of constant $\langle V_{R}\rangle$ in Figure 2e are sketched as dark blue lines in 2f. We also label the index $n$ of the Shapiro step (such that $ V= n\frac{f_{ac}}{k}$). Note that the overall slopes of the plateaus are identical to the contours $\langle V_{L}\rangle=0$ and $\langle V_{R}\rangle=0$ in the DC regime, which are shown in Ref.\cite{suppinfo}. In Figure 2d, the imprint of the $\langle V_{R}\rangle$ plateaus is observed as slightly lighter blue stripes. This is because the sudden drop in $\langle V_{R}\rangle$ causes the effective resistance from the left contact to ground to drop slightly because of the resistor network connecting the junctions. Similarly, plateaus of constant $\langle V_{L}\rangle$ affect $\langle V_{R}\rangle$ and are visible as light blue stripes in Figure 2e. The transverse junction can also become phase-locked. When this happens, $\langle V_{L}-V_{R}\rangle$ is quantized, which forms plateaus parallel to the contour $\langle V_{L}-V_{R}\rangle=0$ in the DC regime.  These correspond to diagonal stripes spanning the map from the bottom left to the top right corner (sketched in pink on Figure 2f). 

Finally, we observe that classical quartet states also yield Shapiro steps, although these are fainter, which are visible on both maps. The most noticeable correspond to plateaus of quantized $\langle V_{L}+V_{R}\rangle$ and quantized $\langle 2V_{L}-V_{R}\rangle$ and are sketched in light blue on Figure 2f. Voltage steps are observed whenever $pV_{l}+qV_{r}=\frac{f_{ac}}{k}$, with (p,q) =(1,1) and (2,-1). For a true Josephson junction network, this would correspond to $pV_{l}+qV_{r}=\frac{hf_{ac}}{2e}$. 

\section{Fractional phase locking}

Maps of multiterminal Shapiro steps are only this simple when the junctions are sufficiently overdamped. Indeed, when the quality factor of the junctions is increased, additional fractional phase locked steps are observed. Fractional Shapiro steps can easily be seen even in single junctions with a sinusoidal CPR\cite{Amet2021}, but can also result from the interaction of two junctions within a network\cite{Arnault2021}. The measurement scheme is identical to what was discussed in the previous section. Figures 3a and 3b show the same differential resistances $\partial \langle V_{L}\rangle/\partial I_{L}$ and $\partial\langle V_{R}\rangle/\partial I_{R}$ for a slightly underdamped analog junction network with a larger quality factor of $\approx$0.8. Plateaus are visible in both channels at integer phase locking for each junction, similar to what was described in Figure 2. However, new plateaus emerge at fractional multiples of the voltage quantum $f_{ac}/k$. These are visible as smaller stripes of vanishing differential resistance which for example can be seen around ($I_{R}=-1.4$mA, $I_L=1.1$mA) for Figure 3a and around ($I_{R}=-1.2$mA, $I_L=1.3$mA) for Figure 3b. To gain insights into those patterns, we plot a cross section of $\langle V\rangle(I_{R})$ along a Shapiro step of $V_{L}$, indicated by a yellow dashed line on panel 3a. We observe robust plateaus in $\langle V_{R}\rangle$ at fractional values with denominators up to 5. Similarly, Figure 3d correspond to a cross section at constant $\langle V_{R}\rangle$ along the full yellow line indicated in panel B. The cross section shows fractional steps in $\langle V_{L}\rangle$

Note that fractional steps in $\langle V_{L}\rangle$ are only observed when $\langle V_{R}\rangle$ is integer phase-locked, the converse being also true. For example the fractions highlighted in Figure 3c are obtained on top of the n=1 plateau of $\langle V_{R}\rangle$. This implies that integer phase-locking of one of the phases tends to stabilize fractions in the other channel.  

We now turn to a time-domain analysis of the phase when fractional Shapiro steps are observed. We first record the unfiltered $V_{L}(t)$ and $V_{R}(t)$ at each bias value, then compute the fast Fourier transform in order to determine their frequency spectrum. We can then generate a map of that frequency spectrum along a bias cross section \cite{Amet2021}. 
Figure 3e shows the spectral weight of the FFT as a function of frequency and bias, when the bias is evolving along the diagonal shown in Panel a. The main resonance in the frequency spectrum is of course the fundamental excitation frequency $f_{ac}$ observed at 687.7 Hz. We see that whenever the I-V curve of the time filtered $\langle V_{L}\rangle$ shows a fractional Shapiro step, the frequency spectrum of the unfiltered $V_{L}(t)$ has stable sub-harmonics at $f_{ac}/q$, where q is the denominator of the fraction. This type of frequency spectrum is reminiscent of what is observed in a single analog junction in the presence of fractional Shapiro steps \cite{Amet2021}.

To understand this behavior, as well the origin of the fractional steps, we turn to numerical simulations of the trajectories in phase space under different bias conditions. Details of those simulations are available in Ref.\cite{suppinfo}. Figure 3e shows a map of the washboard potential, which if we drop a multiplicative constant can be written as $U(\varphi_{L},\varphi_{R})\propto-I_{cL}\sin(\varphi_{L})-I_{cR}\sin(\varphi_{R})-I_{cT}\sin(\varphi_{L}-\varphi_{R})$. We plot two simulated trajectories of the phase over 12 cycles of the drive, and shift them by multiples of $2\pi$ to fit in this window. They correspond to $\varphi_{R}$ in the n=0 phase locked state, while $\varphi_{L}$ is in the n=1/2 (left) or n=1/3 (right) phase locked state. In both cases $\varphi_{R}$ does not drift and just rocks back and forth. In the n=1/2 case, we see that at every other oscillation of the drive, the phase oscillates either across a minimum of the washboard potential, or along the saddle point between two maxima. Those two types of oscillations differ in amplitude, which explains the period doubling of the signal. In the n=1/3 case, the phase oscillates twice along the saddle point and once across a minimum of potential, thus explaining the period tripling observed in panel 3d. In both cases, oscillations around an otherwise unstable saddle point are dynamically stabilized by the rocking of $\varphi_{R}$. While fractional Shapiro steps can be observed in a single current-driven junction, our simulations strongly suggest that the fractional steps we observe are instead caused by the two-dimensional nature of the washboard potential for a three-terminal junction. This mechanism is reminiscent of the dynamical stabilization of classical multiplet supercurrents observed in Ref.\cite{arnault_dynamical_2022}, which was shown to be mathematically equivalent to Kapitza's inverted pendulum problem. This also explains the need for a larger quality factor to observe such fractions, since inertia facilitates this stabilization\cite{arnault_dynamical_2022}. 

\begin{figure}
    \centering
    \includegraphics[width=0.47\textwidth]{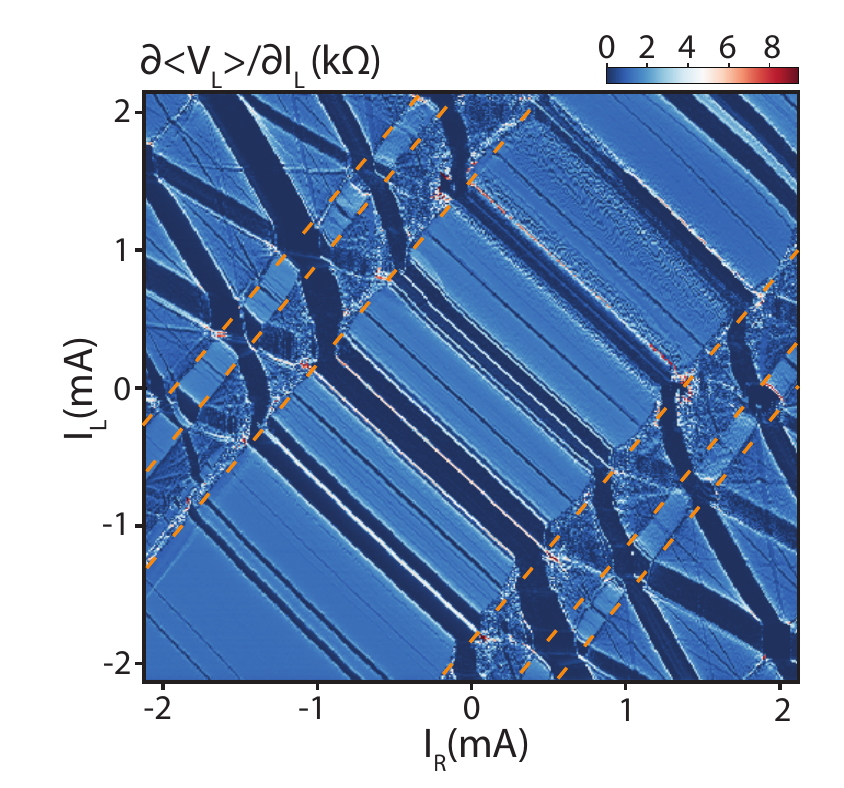}
    \caption{Differential resistance $\partial\langle V_{L}\rangle/\partial I_{L}$ as a function of both biases $I_{L}$ and $I_{R}$. The map was measured with $f_{ac}=286.7 Hz$ and $I_{ac}$=1.8 mA. The effective transverse critical current was increased by a factor 2.5 for this measurement, by dropping $R_{T2}$ from 7.5 $k\Omega$ to 3 $k\Omega$}
    \label{fig:mjj_diagram}
\end{figure}

\section{Synchronization}

Finally, we turn to the impact of transverse coupling on the synchronization of the two phases. To that end, we increase the value of the transverse coupling $I_{cT}$, which can just be done by reducing $R_{T2}$. We then measure the differential resistance $\partial \langle V_{L}\rangle/\partial I_{L}$, which is shown on Figure 4. The map of $\partial \langle V_{R}\rangle/\partial I_{R}$ is essentially identical.

We focus in this section on the three stripes of suppressed differential resistance going from the bottom left to top right corners of the map, and whose boundaries are highlighted with orange dashed lines. These correspond to the quantization of the voltage across the transverse junction with $n=-1, 0, 1$. In those regions, the stability contours of the plateaus of $\langle V_{L}\rangle$ and $\langle V_{R}\rangle$ are identical because the two voltages are locked to each other by the quantization of the transverse junction. This explains why the slope of those plateaus changes as they intersect with plateaus of the transverse junction. We can also see that when the three types of stripes intersect, all three junctions are phase locked to an integer multiple of $f_{ac}/k$. This occurs for example around $I_{L}=1$mA, $I_{R}=-1$mA, which corresponds to quantized voltages $\langle V_{L}\rangle =-2f_{ac}/k$, $\langle V_{r}\rangle =-f_{ac}/k$ and $\langle V_{L}-V_{R}\rangle =-f_{ac}/k$. 

Note that in a conventional Josephson junction a voltage quantum is $hf_{ac}/2e$, whereas in the analog equivalent it is $f_{ac}/k$, where k is the voltage to frequency gain of the junction. It is therefore important to calibrate the gain of the three VCOs so they are as close to each other as possible. We fine tuned them so that $k_{L}=1817 Hz/V$, $k_{R}=1818 Hz/V$ and $k_{T}=1818 Hz/V$. Despite this calibration, we observe some artifacts that are caused by the non universal size of Shapiro steps. These are most visible on top  of the n=0 plateau of the transverse junction, when it intersects plateaus of the other two junctions. These correspond to dark blue bands of slope $\approx -1$ perpendicular to the widest plateau of slope $\approx 1$. We see that each plateau splits into two in that region, which is barely noticeable in a voltage map, but striking in a differential resistance map.

Our results provide a convenient table-top alternative to pure numerical modeling to observe and classify dynamical phenomena in multiterminal junctions. The analog platform makes it possible to vary every RCSJ circuit parameter, and better understand the role of transverse coupling and quality factor in the behavior of a junction network. It will facilitate the interpretation of transport measurements on real multiterminal Josephson junctions and make it possible to single-out quantum effects from dynamical ones. 

\section{Acknowledgements}

We thank Wade Hernandez, Trevyn Larson and Patrick Richardson for their input and technical help. S.I. was supported by a GRAM fellowship. F.A., and A. M. were supported by a URC grant at Appalachian State University. E.G.A. was supported by the Division of Materials Sciences and Engineering, Office of Basic Energy Sciences, U.S. Department of Energy, under Award No. DE-SC0002765. 

\providecommand{\noopsort}[1]{}\providecommand{\singleletter}[1]{#1}%

\clearpage

\onecolumngrid
\renewcommand{\thepage}{S\arabic{page}}
\renewcommand{\thefigure}{S\arabic{figure}}
\renewcommand{\thesection}{S\arabic{section}}
\setcounter{figure}{0}
\setcounter{page}{1}
\setcounter{equation}{0}
\setcounter{section}{0}

\begin{center}
    {\LARGE \textbf{Supplementary information}}
\end{center}

\begin{figure}[b]
    \centering
    \includegraphics[width=\textwidth]{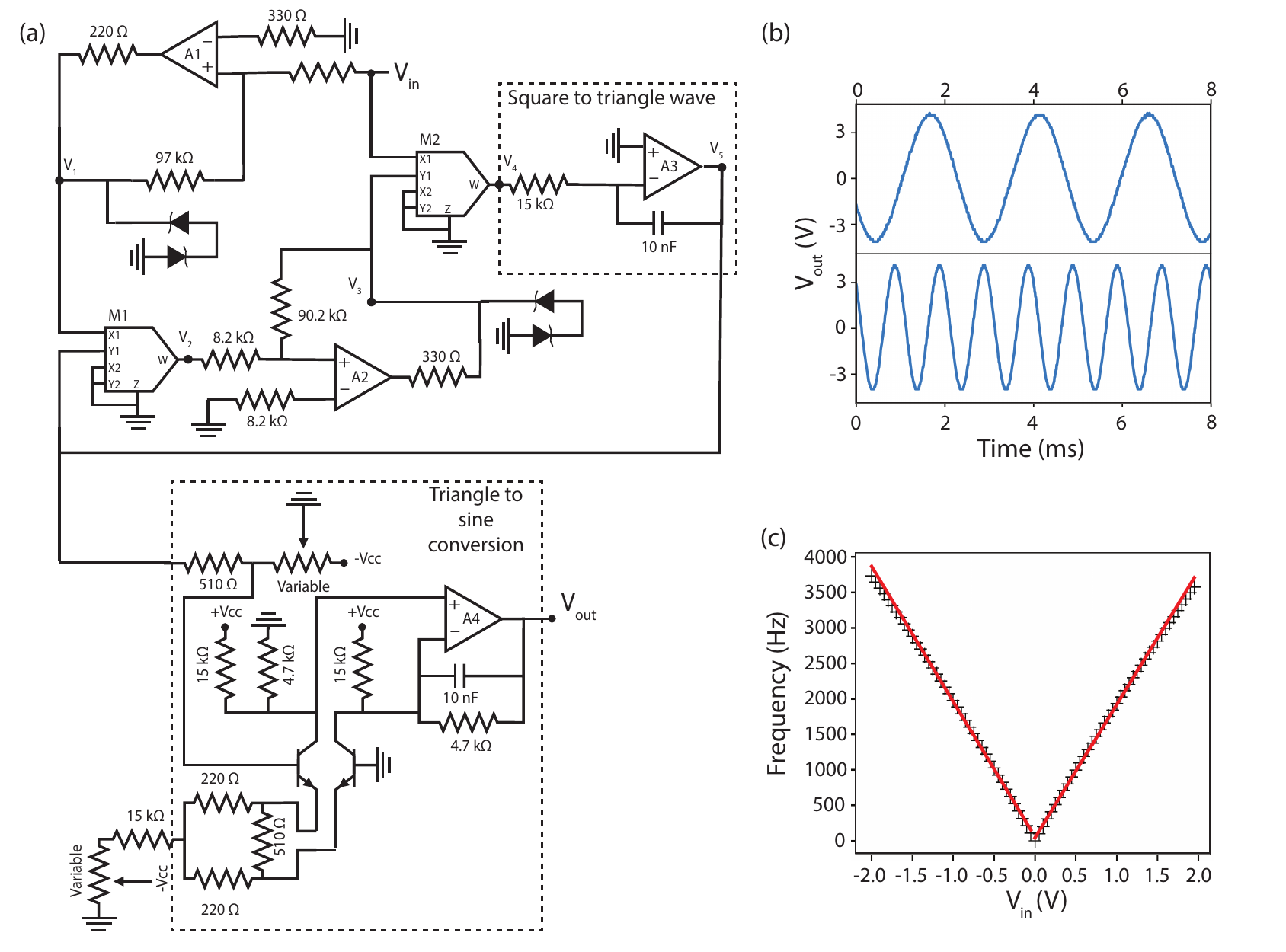}
    \caption{(a) Circuit diagram of one of the three voltage controlled oscillators used in the main circuit. (b) The output voltage of the VCO is sinusoidal with a frequency that is proportional to the input voltage. It is shown here when the input voltage is 0.2V (top) and 0.5V (bottom). (c) Output frequency of the VCO as a function of the input voltage. The voltage to frequency gain is approximately 1800 Hz/V.}
    \label{fig:vco_figure}
\end{figure}
\twocolumngrid

\vspace{20mm}

\section{Characterization of the voltage controlled oscillators}

Figure S1a is a diagram of one of the voltage controlled oscillators used in this paper. The part of the circuit that generates the triangle wave was proposed by Ref.\cite{blackburn_circuit_2007}. An explanation of the behavior of this circuit is provided in Ref.\cite{Amet2021}. It outputs a sine wave whose frequency is proportional to the input of the VCO ($V_{in}$). Figure S1b shows two examples of output sine waves for two different DC inputs (0.2V for the top graph, 0.5V for the bottom one). We demonstrated in Ref. \cite{Amet2021} that the triangle to sine converter in the circuit heavily suppressed higher harmonics. Figure S1c shows the frequency of the output as a function of the input voltage, which is what is used to determine the voltage to frequency gain, on the order of 1800 Hz per volt. 
\begin{figure*}
    \centering
    \includegraphics[width=0.99\textwidth]{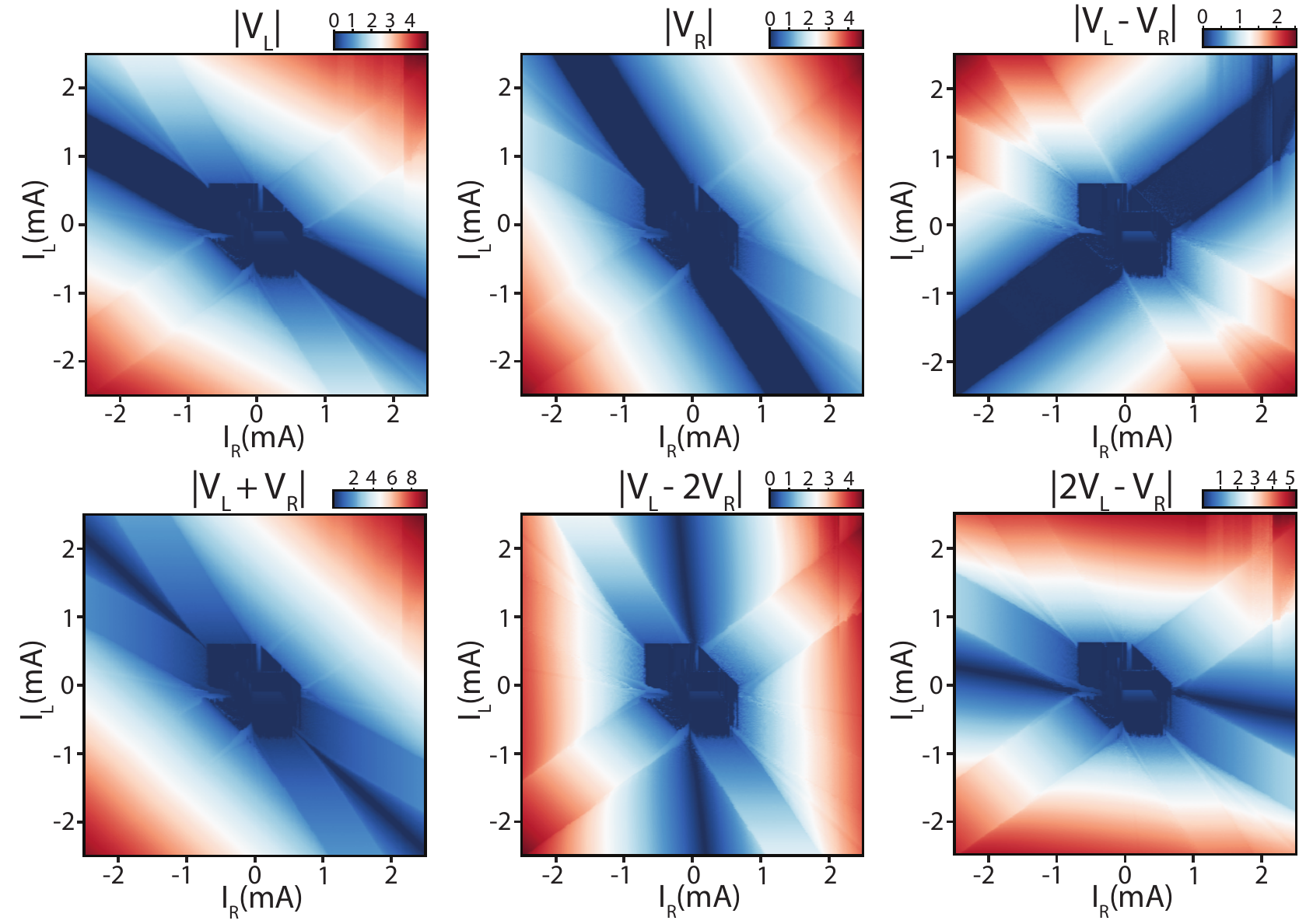}
    \caption{(a-c)Maps of the voltages $\langle V_{L}\rangle$, $\langle V_{R}\rangle$ and $\langle V_{L}-V_{R}\rangle$ as a function of both biases $I_{L}$ and $I_{R}$ with no AC drive. (d-f)Maps of the voltages $\langle V_{L}+V_{R}\rangle$, $\langle V_{L}-2V_{R}\rangle$ and $\langle 2V_{L}-V_{R}\rangle$. Shapiro steps boundaries that are parallel to these contours are signatures of the inverse AC Josephson effect of quartet states.}
    \label{fig:S2}
\end{figure*}

\begin{figure*}
    \centering
    \includegraphics[width=0.99\textwidth]{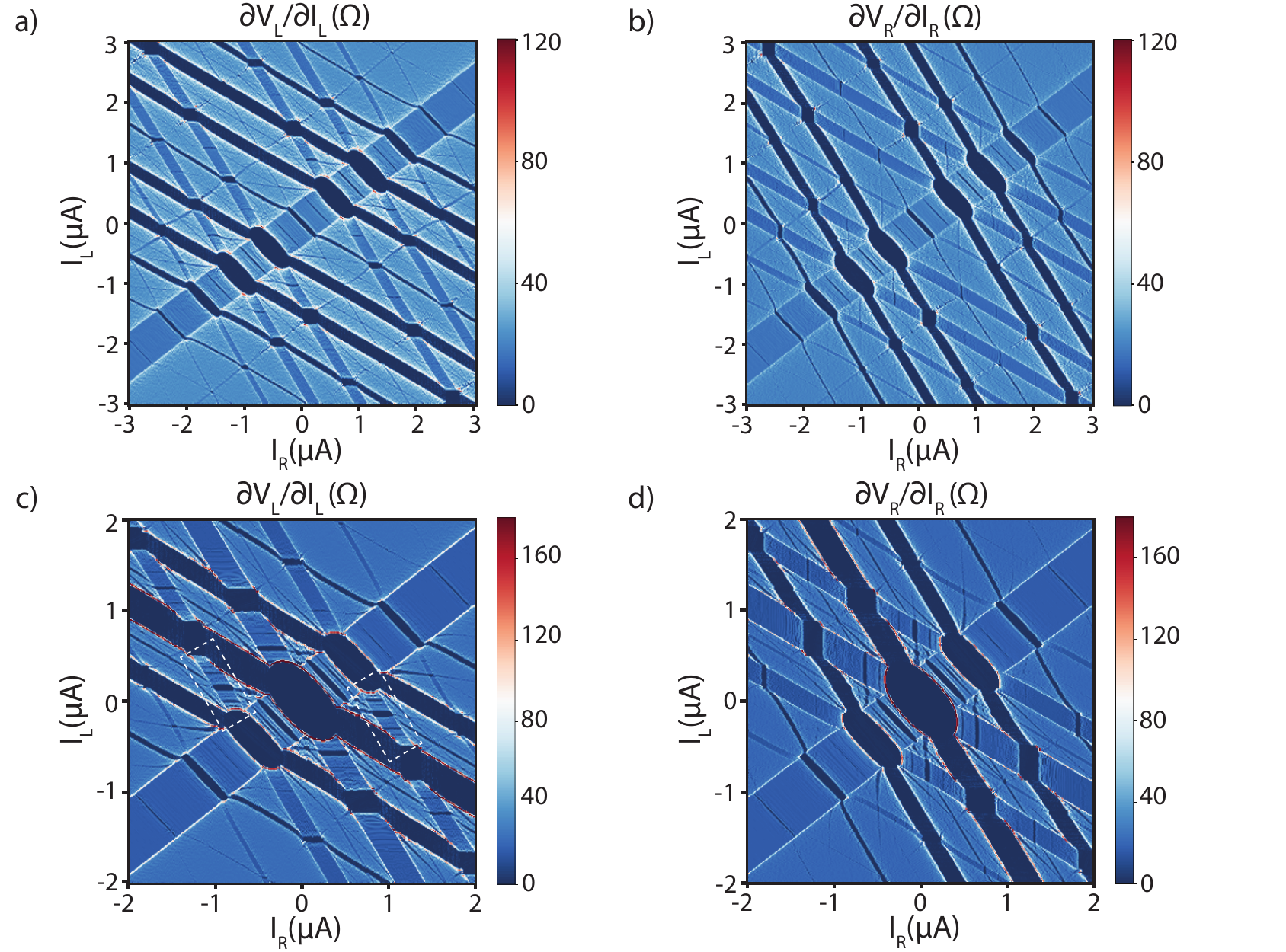}
    \caption{a-b) Numerical simulations of the differential resistances $\partial \langle V_{L}\rangle/\partial I_{L}$ and $\partial \langle V_{R}\rangle/\partial I_{R}$ as a function of both biases $I_{L}$ and $I_{R}$. Those two simulations are ran for an average quality factor of 0.5, $I_{ac}=1.7$$\mu$A $f_{ac}=8$GHz. c-d) Simulated differential resistances in a different regime where Q=1.3, and $I_{ac}=2.8$$\mu$A (the other circuit parameters being unchanged). Fractional Shapiro steps are now much more pronounced, for example in regions boxed with white dashed lines.}
    \label{fig:S3}
\end{figure*}

Note that in this work, the gains of the VCOs must be closely adjusted so they yield Shapiro steps of equal voltage amplitudes, since those are equal to $f_{ac}/k$. This is of course in sharp contrast to a real Josephson junction network where Shapiro steps would reliably be equal to $hf_{ac}/2e$. The gain $k$ is best controlled by the $\approx 15$k resistor which is part of the square to triangle wave converter, and which is a potentiometer. Slightly changing the value of that resistor changes the gain of the VCO without altering the amplitude of the oscillation. This allows us to tune $k$ for all VCOs within $0.2\%$ of each other. Despite those adjustments, the differential maps presented in the main paper still present minor artifacts that result from the slight difference in Shapiro step sizes across the three junctions.  

\section{Derivation of the phase dynamics}

We define $\dot{\varphi}_{L}\equiv 2\pi k V_{L} $ and $\dot{\varphi}_{R}\equiv 2\pi k V_{R}$, where $V_{L}$ and $V_{R}$ are the voltage outputs of amplifiers A3 and A2 in Figure 1a of the main paper. With those notations, the voltage at the node between the left junction's VCO and $R_{L2}$ is  $\alpha\sin(\varphi_{L})$, whereas the voltage at the node between the right junction's VCO and $R_{R2}$ is  $\alpha\sin(\varphi_{R})$.

We call $I_{T}$ the current flowing through the bottom horizontal branch of the circuit diagram.

In this circuit, A4 is used as a differential amplifier, A1 is a current to voltage converter, and A5 is a unity-gain voltage inverter. Assuming that the amplifiers have a vanishing current input, we apply Kirchhoff rules at the inverting inputs of A2 and A3 and find:
\begin{equation}
    -I_{T}+I_{L}+\frac{\alpha}{R_{L2}}\sin(\varphi_{L})+\frac{\dot{\varphi_{L}}}{2\pi k R_{L}}+C_{L}\frac{\ddot{\varphi_{L}}}{2\pi k}=0
\end{equation}

\begin{equation}
    I_{T}+I_{R}+\frac{\alpha}{R_{R2}}\sin(\varphi_{R})+\frac{\dot{\varphi_{R}}}{2\pi k R_{R}}+C_{R}\frac{\ddot{\varphi_{R}}}{2\pi k}=0
\end{equation}

A4 generates a voltage $(\dot{\varphi_{L}}-\dot{\varphi_{R}})/(2\pi\alpha k)$, which is then fed to the subcircuit emulating the transverse junction. The voltage at the node between the transverse junction VCO and $R_{T2}$ is therefore $\alpha\sin(\varphi_{L}-\varphi_{R})$. We find that:
\begin{equation}
I_{T}=\frac{\alpha}{R_{T2}}\sin(\varphi_{L}-\varphi_{R})+\frac{\dot{\varphi_{L}}-\dot{\varphi_{R}}}{2\pi k R_{T}}+C_{T}\frac{\ddot{\varphi_{L}}-\ddot{\varphi_{R}}}{2\pi k}
\end{equation}

Using the same matrix notation as before, we finally get:
\begin{align}
    \frac{1}{2\pi k}\mathcal{C} \ddot\Phi +\frac{1}{2\pi k}\mathcal{G} \dot\Phi+I_{c}(\Phi)=I
\end{align}

Where we defined $I_{cL}=\frac{\alpha}{R_{L2}}$, $I_{cT}=\frac{\alpha}{R_{T2}}$, and $I_{cR}=\frac{\alpha}{R_{R2}}$, and:  

\begin{align*}
I_{c}(\Phi)&=\begin{pmatrix}I_{cL}\sin(\varphi_{L})+I_{cT}\sin(\varphi_{L}-\varphi_{R})\\
    I_{cR}\sin(\varphi_{R})+I_{cT}\sin(\varphi_{R}-\varphi_{L})
    \end{pmatrix}
\end{align*}

We thus recover the same system of differential equations as for a three terminal shunted Josephson junction network, where the constant $\hbar/2e$ was replaced by $1/(2\pi k)$. Note that in this system, the effective critical currents $I_{cL}$, $I_{cR}$ and $I_{cT}$ can be tuned by changing the resistances $R_{L2}$, $R_{R2}$ and $R_{T2}$ respectively.

\section{Additional DC characterization}

Panels a through c of Figure S2 are maps of the three main junction voltages $\langle V_{L}\rangle$, $\langle V_{R}\rangle$ and $\langle V_{L}-V_{R}\rangle$. Contours of vanishing voltages correspond to the three resonances of vanishing effective differential resistance seen in Figure 1b and 1c of the main paper. This behavior is identical to what was observed in typical three-terminal Josephson junction transport measurements \cite{Draelos2019,Arnault2021, Pankratova2020,Pribiag2020}.

Panels d through f of Figure S2 show the bias dependence of $\langle V_{L}+V_{R}\rangle$, $\langle V_{L}-2V_{R}\rangle$ and $\langle 2V_{L}-V_{R}\rangle$. When those voltages are zero, a narrow resonance of vanishing differential resistance is observed in  the DC transport maps of Figure 1. These resonances correspond to the three possible types of classical quartet 'supercurrents'. Ref. \cite{arnault_dynamical_2022} studies in greater detail the dynamical origin of those quartet resonances. Note that in Figures 2 and 3, we also observe Shapiro steps parallel to those contours, which correspond to the phase locking of quartet states.  

\section{Numerical simulation of Inverse AC Shapiro}

As discussed in the main paper, to describe the time evolution of the phase in a three-terminal Josephson junction, one can define $\Phi=\begin{pmatrix} \varphi_{L} \\ \varphi_{R} \end{pmatrix}$ and get:

\begin{equation}
\frac{\hbar}{2e}\mathcal{C} \ddot\Phi +\frac{\hbar}{2e}\mathcal{G}\dot\Phi + I_c(\Phi)=I
\end{equation}

That system of differential equations can be first turned into a four-dimensional first order equation:

\begin{align*}
    \dot{\Phi}&=\Psi\\
    \dot{\Psi}&=-\mathcal{C}^{-1}\mathcal{G}\Psi+\frac{2e}{\hbar}\mathcal{C}^{-1}(I-I_c(\Phi))
\end{align*}

That system of equation can then be solved by a fourth order Runge-Kutta routine, which yields the time dependence of $\Phi(t)$ for a given bias. The voltages $V_{L}$ and $V_{R}$ across the left and right junctions can then be obtained by computing $\hbar\langle\dot{\Phi}\rangle/2e$ over a few cycles of the drive. Note, however, that rather than solving the time evolution of the phases for each bias values sequentially, we implement a vectorized code where phase dynamics at all bias values are solved in parallel, which speeds up the simulation by two orders of magnitude. Additional details on this technique are described in Ref.\cite{Arnault2021}. 

We did not attempt to perform a quantitative fit between our data and this model, which would be impractical due to the large size of the parameter space (the circuit network includes 11 fitting parameters). However, we can reproduce qualitatively the behavior observed on the analog circuit. Figures S3a and S3b show the differential resistances $\partial V_{L}/\partial I_{L}$ and $\partial V_{R}/\partial I_{R}$ in an overdamped regime (Q $\approx$0.2) that replicates patterns observed in Figure 2 of the main paper. In particular, integer Shapiro steps corresponding to all three junctions, as well as some multiplet resonances, can be seen. Figures S3c and S3d show the same differential resistances for a junction with a larger average quality factor of approximately 1.5. Additional, more robust fractional Shapiro steps are observed, in qualitative agreement with Figures 3a and 3b of the main paper. A table with the circuit parameters used in both simulations is provided below:

\begin{center}
\begin{tabular}{|c| c |c |} 
 \hline
 Parameters & Simulation 1 & Simulation 2 \\ 
 \hline
 $f_{ac}$ & 8 GHz & 8 GHz  \\ 
 \hline
 $I_{ac}$ & 1.7 $\mu$ A & 2.8 $\mu$ A\\ 
 \hline
 $C_{L}=C_{R}=C_{T}$ & 50 fF & 3000 fF  \\ 
 \hline
 $R_{L}$ & 39 $\Omega$ & 39 $\Omega$ \\ 
 \hline
 $R_{R}$ & 29 $\Omega$ & 29 $\Omega$ \\
 \hline
 $R_{T}$ & 24 $\Omega$ & 24 $\Omega$ \\
 \hline
 $I_{L}$ & 275 nA & 275 nA \\
 \hline
 $I_{R}$ & 250 nA & 250 nA \\
 \hline
 $I_{T}$ & 245 nA & 245 nA \\

 \hline
\end{tabular}
\end{center}





\end{document}